\documentclass[onecollarge,natbib]{svjour2}
\bibpunct{[}{]}{;}{n}{}{,} 
\smartqed  
\usepackage{graphicx}
\usepackage{amsmath}
\journalname{Few-Body Systems (EFB22)}
\begin{document}

\title{Multicomponent Strongly Interacting Few-Fermion
Systems in One Dimension
}


\author{Artem~G.~Volosniev \and Dmitri~V.~Fedorov \and Aksel~S.~Jensen \and Nikolaj~T.~Zinner \and Manuel~Valiente       
}


\institute{Artem~G.~Volosniev \and Dmitri~V.~Fedorov \and Aksel~S.~Jensen \and Nikolaj~T.~Zinner \at
              Department of Physics and Astronomy, Aarhus University, DK-8000 Aarhus C, Denmark           
           \and
           Manuel~Valiente \at SUPA, Institute for Photonics and Quantum Sciences,
Heriot-Watt University, Edinburgh EH14 4AS, UK   
 \email{artem@phys.au.dk}
}


\date{Received: date / Accepted: date}

\maketitle

\begin{abstract}
The paper examines a trapped one-dimensional system of multicomponent spinless fermions
that interact with a zero-range two-body potential. We show that when
the repulsion between particles is very large the system can be approached
analytically. To illustrate this analytical approach 
we consider a simple system of three distinguishable particles, which can be addressed experimentally.
For this system we show that for infinite repulsion the energy spectrum is six fold degenerate. 
We also show that this degeneracy is partially lifted for finitely large repulsion for which 
we find and describe corresponding wave functions. 

\end{abstract}

\section{Introduction}
\label{intro}

Experimental study of few-body physics in cold atomic gases is 
a complicated task since 
such systems usually have a relatively large particle density 
and, hence, many-body correlations should be taken into account. 
Only very recently setups with small particle numbers 
were realized in Heidelberg \cite{serwane_thesis, serwane,zurn} where ground state systems
of a few fermionic atoms, $^6Li$, were prepared in a quasi-one-dimensional trap. 
Such setups pave the way 
for the experimental study of few-body correlations where accurate 
theoretical description can be obtained through advanced numerical investigation \cite{lindgren,gharashi}. 
This paper overviews the newly developed method \cite{volosniev} to study the mentioned experimental setups
in the limit of strong repulsion between particles without 
applying complicated numerical routines. 
This analytical approach gives a description of the Tonks-Girardeau gas \cite{tonks, girardeau,
lieb} of a few particles in a trap.  Moreover, as a by-product results obtained with 
this method can be used as a reference point for numerical calculations.

The structure of the paper is the following: in section~\ref{sec:1} we  introduce the Hamiltonian that is widely 
used 
to describe the relevant experimental setups \cite{serwane_thesis,zurn,lindgren,gharashi,volosniev,sowinski,cui}, 
in section~\ref{sec:2} we illustrate the approach using the 
simple system of three distinguishable particles in a harmonic trap, that to the best of 
our knowledge was not addressed before. There we find and describe 
eigenstates of such system in the limit of strong interparticle interaction.

\section{Formulation of the problem}
\label{sec:1}

This paper considers $N$ particles of equal mass $m$ in one spatial dimension. 
Additionally, the following assumptions are applied:
$i)$ the particles can be divided into classes of identical spinless fermions;
$ii)$ the system is trapped by some external potential, $V_{ext}$;
$iii)$ the interparticle interaction is assumed to be of zero range, $V=g\delta(x_i-x_j)$, where 
$g$ is a strength parameter and $x_i,x_j$ are the coordinates of particles $i$ and $j$.
These assumptions lead to the following Hamiltonian 
\begin{equation}
\label{hamilt}
H=-\frac{\hbar^2}{2m}\sum_{i=1}^N\frac{\partial^2}{\partial x_i^2}+\sum_{i=1}^NV_{ext}(x_i)+g\sum_{i>j}\delta(x_i-x_j).
\end{equation}
This model Hamiltonian is often used to study the relevant experimental setup
of a few fermionic atoms $^6Li$ in different hyperfine states 
\cite{serwane_thesis,zurn,lindgren,gharashi,volosniev,sowinski,cui}. This 
model Hamiltonian allows one to explore analytical approach only in limiting cases, e.g.
two particles in a harmonic trap~\cite{busch} or $N$ particles without an external confinement~\cite{McGuire},
such that the theoretical study for the relevant experiments 
is usually provided using different numerical techniques, e.g. \cite{lindgren, gharashi}. 
It is shown in \cite{volosniev} that one can find eigenstates for such Hamiltonian 
in the case of very strong repulsion between particles, i.e. if $1/g\to 0$. 
This regime is experimentally accessible \cite{zurn} and very interesting 
from the theoretical point of view since it allows a theoretician to gain knowledge about strongly interacting systems.

In this paper we overview 
the method from \cite{volosniev} using a simple system of 
three distinguishable particles trapped in a harmonic oscillator trap.
In other words we consider the Schr{\"o}dinger equation 
\begin{equation}
\left(-\frac{\hbar^2}{2m}\sum_{i=1}^3\frac{\partial^2}{\partial x_i^2}+\frac{m\omega^2}{2}\sum_{i=1}^3x_i^2+g\delta(x_1-x_2)+g\delta(x_2-x_3)+g\delta(x_1-x_3)\right)\Psi=E\Psi,
\label{schr-eq}
\end{equation}
where $\Psi$ is the wave function, $E$ is the energy of the system, $\omega$ is the frequency of the harmonic oscillator, 
$g$ is assumed to be large and positive, such that $1/g\rightarrow0$.
Eq.~(\ref{schr-eq}) can be recast into the free Schr{\"o}dinger equation
plus the boundary conditions
\begin{equation} 
\label{eq-ferm-cond}
\left(\frac{\partial\Psi }{\partial x_i}-\frac{\partial\Psi }{\partial x_j}\right)\bigg|_{x_i-x_j=+0}-
\left(\frac{\partial\Psi }{\partial x_i}-\frac{\partial\Psi }{\partial x_j}\right)\bigg|_{x_i-x_j=-0}=\frac{2gm}{\hbar^2}\Psi\bigg|_{x_i=x_j},
\end{equation}
where $i,j=1,2,3$ and $i\neq j$. Eqs.~(\ref{schr-eq}) and (\ref{eq-ferm-cond}) 
contain all ingredients  that are needed to describe a general method \cite{volosniev}; moreover, 
the system of three distinguishable particles is interesting 
on its own rights as experimentally relevant. 

\section{Method}
\label{sec:2}

\subsection{Case of infinite repulsion between particles, $1/g=0$.}
To solve eq.~(\ref{schr-eq}) for large values of $g$  
we first notice that if $1/g=0$, then eq.~(\ref{eq-ferm-cond}) yields $\Psi|_{x_i=x_j}=0$,
which means that eq.~(\ref{schr-eq}) can be seen as an equation for free particles 
that cannot penetrate through one another. This problem formally resembles the case 
of three identical spinless fermions with the only difference that now the total wave function 
not need to have a continuous derivative when two particles meet.
This possibility to have a discontinuous 
derivative follows from eq.~(\ref{eq-ferm-cond}). 
However, since the particles cannot penetrate through one another it is enough to solve 
the Schr{\"o}dinger equation only for a given ordering of the particles, i.e. $x_1<x_2<x_3$, with the boundary conditions
$\Psi_{x_1=x_2}=\Psi_{x_2=x_3}=0$. These boundary conditions can be satisfied 
only for the energies from the eigenspectrum of spinless fermions $\{E_0,E_1,...\}$, which implies that 
the total wave function for $1/g=0$ can be built using the wave function of spinless fermions $\Psi_F$
for a given energy $E_i$ 
\begin{equation}
\label{eq-ferm-wave}
\Psi=\left\{
  \begin{array}{l l}
     a_1 \Psi_F \; {\rm for} \; x_2<x_1<x_3 \;  \\
    a_2 \Psi_F  \; {\rm for} \;  x_2<x_3<x_1 \; \\
	 a_3\Psi_F \; {\rm for} \; x_3<x_2<x_1 \;   \\
	   a_4 \Psi_F \; {\rm for} \; x_3<x_1<x_2 \\
	 a_5 \Psi_F \; {\rm for} \; x_1<x_3<x_2 \;   \\
	  a_6 \Psi_F \; {\rm for} \; x_1<x_2<x_3   \; 
  \end{array} \right.
\end{equation}
where $a_i$ are real coefficients. To obtain the wave function~(\ref{eq-ferm-wave})
we assumed that the energy $E_i$ corresponds to only one wave function $\Psi_F$. 
The extension of the method for the more general situation when the energy $E_i$ corresponds
to more than one wave function is discussed in refs.~\cite{volosniev, vol_thesis}.  The possibility 
to build the wave function in the form of eq.~(\ref{eq-ferm-wave})
implies that for $1/g=0$ the energy spectrum of eq.~(\ref{schr-eq}) is six fold degenerate,
since all six coefficients $a_i$ in eq.~(\ref{eq-ferm-wave}) are linearly independent.   
\subsection{Case of finitely large repulsion between particles, $1/g\rightarrow0$.}
{\it Method}. To find the behavior of the energy in the vicinity of $1/g=0$  we use the Hellmann-Feynman theorem 
to obtain the derivative of the energy 
\begin{equation}
\frac{\partial E}{\partial g} = \frac{\int \mathrm{d}x_1\mathrm{d}x_2 (\Psi)^2_{x_3=x_1}+\int \mathrm{d}x_1\mathrm{d}x_3 (\Psi)^2_{x_2=x_3}+\int \mathrm{d}x_2\mathrm{d}x_3 (\Psi)^2_{x_1=x_2}}{\langle \Psi|\Psi \rangle}\; .
\end{equation}
Next we notice that from eq.~(\ref{eq-ferm-cond}) it follows that
$\Psi|_{x_i=x_j}\sim1/g+o(1/g)$, which allows us to conclude 
that $E\sim E_i - K/g+o(1/g)$, where the parameter $K$
is given by 
\begin{equation}\label{de}
K\doteq\lim_{g\to\infty}g^2\frac{\partial E}{\partial g}=\lim_{g\to\infty}\,g^2\frac{\int \mathrm{d}x_1\mathrm{d}x_2 (\Psi)^2_{x_3=x_1}+\int \mathrm{d}x_1\mathrm{d}x_3 (\Psi)^2_{x_2=x_3}+\int \mathrm{d}x_2\mathrm{d}x_3 (\Psi)^2_{x_1=x_2}}{\langle \Psi|\Psi\rangle}.
\end{equation}
The limit in this equation can be easily taken, since $(\Psi)^2_{x_i=x_j}\sim1/g^2+o(1/g^2)$, which yields 
\begin{align}\label{eq-ferm-K}
K=\dfrac{\hbar^4}{m^2}\dfrac{\int \big(\frac{1}{2}(\frac{\partial }{\partial x_1}-\frac{\partial }{\partial x_3})|_{x_3-x_1=+0}\Psi-
\frac{1}{2}(\frac{\partial }{\partial x_1}-\frac{\partial }{\partial x_3})|_{x_3-x_1=-0}\Psi\big)^2\mathrm{d}x_1\mathrm{d}x_2}{\langle \Psi|\Psi\rangle} +\\ \nonumber
\frac{\hbar^4}{m^2}\frac{\int \big(\frac{1}{2}(\frac{\partial }{\partial x_2}-\frac{\partial }{\partial x_3})|_{x_2-x_3=+0}\Psi-
\frac{1}{2}(\frac{\partial }{\partial x_2}-\frac{\partial }{\partial x_3})|_{x_2-x_3=-0}\Psi\big)^2\mathrm{d}x_1\mathrm{d}x_3}{\langle \Psi|\Psi\rangle}+\\ \nonumber
\frac{\hbar^4}{m^2}\frac{\int \big(\frac{1}{2}(\frac{\partial }{\partial x_1}-\frac{\partial }{\partial x_2})|_{x_1-x_2=+0}\Psi-
\frac{1}{2}(\frac{\partial }{\partial x_1}-\frac{\partial }{\partial x_2})|_{x_1-x_2=-0}\Psi\big)^2\mathrm{d}x_2\mathrm{d}x_3}{\langle \Psi|\Psi\rangle}.
\end{align}
Using the wave function (\ref{eq-ferm-wave}) the value of $K$ can be written as
\begin{equation}
K=\gamma\frac{(a_1-a_2)^2+(a_2-a_3)^2+(a_3-a_4)^2+(a_4-a_5)^2+(a_5-a_6)^2+(a_1-a_6)^2}{a_{1}^{2}+a_{2}^{2}+a_{3}^{2}+a_{4}^{2}+a_{5}^{2}+a_{6}^{2}}\; , 
\end{equation}
where $\gamma$ is defined as 
\begin{equation}
\gamma=\dfrac{\hbar^4}{m^2}\dfrac{\int_{x_2<x_3} (\frac{\partial }{\partial x_1}\Psi_F)^2|_{x_1=x_3}\mathrm{d}x_2\mathrm{d}x_3}{\int_{x_1<x_2<x_3\Psi_F^2\mathrm{d}x_1\mathrm{d}x_2\mathrm{d}x_3}}.
\end{equation}
To determine the behavior of the energy we need to find 
the coefficient $K$. To do so we use the variational treatment, i.e. we minimize the energy
by varying $K$ with respect to the coefficients~$a_i$. 
\begin{equation}
\label{eq-system}
\frac{\partial K}{\partial a_i}=0 \; .
\end{equation}
In this way we find the wave functions 
(defined by $a_i$ and eq.~(\ref{eq-ferm-wave})) 
to which the states outside of the degenerate point ($1/g=0$)
are adiabatically connected. This procedure is variational, since it relies on varying $K$
with respect to the coefficients $a_i$. However, it produces an
exact solution, since the wave function 
$\Psi$ for $1/g=0$ can always be written in the form (\ref{eq-ferm-wave}),
so we vary in the full space. 

{\it Solution}.
Equation~(\ref{eq-system}) produces the system of linear equations
\begin{align}
\label{eq-sys} 
-a_{6}+2a_1-a_{2}&=Ka_1/\gamma\; , \qquad \nonumber \\
-a_{i-1}+2a_i-a_{i+1}&=Ka_i/\gamma, \; \mathrm{for} \; i=2,3,4,5 \\
-a_{5}+2a_6-a_{1}&=Ka_6/\gamma\; . \nonumber 
\end{align}
This can be seen as a usual eigenvalue problem for a  
system of linear equations, which can be easily solved using standard techniques 
from linear algebra. The solution contains six eigenstates:
\begin{align}
K_1&=4\gamma  \to a_2=-a_1, a_3=a_1, a_4=-a_1, a_5=a_1, a_6=-a_1 \nonumber \\
K_2&=3\gamma  \to a_2=-a_1/2, a_3=-a_1/2, a_4=a_1, a_5=-a_1/2, a_6=-a_1/2 \nonumber \\
K_3&=3\gamma  \to a_2=-2a_1, a_3=a_1, a_4=a_1, a_5=-2a_1, a_6=a_1 \nonumber \\
K_4&=\gamma  \; \to a_2=a_1, a_3=0, a_4=-a_1, a_5=-a_1, a_6=0 \nonumber \\
K_5&=\gamma  \; \to a_2=0, a_3=-a_1, a_4=-a_1, a_5=0, a_6=a_1 \nonumber \\
K_6&=0  \; \to a_2=a_1, a_3=a_1, a_4=a_1, a_5=a_1, a_6=a_1 \nonumber 
\end{align}
where the coefficient $a_1$ is left for the overall normalization. 
First of all one notices that the lowest energy state defined by $K_1$ has a fully symmetric 
wave function and can be obtained from the Fermi-Bose 
mapping, which is first discussed in ref.~\cite{girardeau}. 
This is always the case for the Hamiltonians in the form of eq.~(\ref{hamilt}),
without any symmetry requirements.
The state with $K_6=0$ corresponds to a fully antisymmetric wave function 
of the system of spinless fermions. Four other states are the same as for the 
system of two spinless fermions and a third particle \cite{lindgren,gharashi,volosniev}.  
The degeneracy of the spectrum, $K_2=K_3$ and $K_4=K_5$, 
comes from the observation that one can antisymmetrize 
either $x_1$ and $x_2$ particles or $x_1$ and $x_3$ particles. 
Combinations of states with $K_2,K_3,K_4,K_5$ also appear for two spinless bosons 
and a third particle as shown in~\cite{zinner}. 

In this way we obtained the total energy spectrum of the system of 
three strongly interacting particles, where six states can be divided 
into three classes: three spinless bosons, three spinless fermions, 
and two spinless fermions interacting with a third particle. 

\section{Conclusions}
\label{sec:3}

In this paper one-dimensional strongly interacting systems of particles were 
investigated. We presented the method to treat such systems 
with zero-range repulsive interactions of large strength. 
To illustrate the approach we analyzed a simple system of three particles.
For such a system we found the energy spectrum and the corresponding 
wave functions.


\end{document}